# Hierarchical community detection via maximum entropy partitions and the renormalization group

Jorge Martinez Armas[a]

[a]Department of Developmental Biology & Neuroscience, Washington University in St. Louis School of Medicine, St. Louis, MO 63110, USA

[1]To whom correspondence may be addressed. Email: mjorge@wustl.edu

ORCID:
J.M.A.: https://orcid.org/0000-0002-9758-967X

**This PDF file includes:**
  Main Text (3938 words)
  Figures 1 to 5
  Methods (2368)
  References (44)

## Abstract

Identifying meaningful structure across multiple scales remains a central challenge in network science. We introduce *Hierarchical Clustering Entropy* (HCE), a general and model-agnostic framework for detecting informative levels in hierarchical community structures. Unlike existing approaches, HCE operates directly on dendrograms without relying on edge-level statistics. It selects resolution levels that maximize a principled trade-off between the entropy of the community size distribution and the number of communities, corresponding to scales of high structural heterogeneity. This criterion applies to dendrograms produced by a wide range of clustering algorithms and distance metrics, including modularity-based and correlation-based methods. We evaluate HCE on synthetic benchmarks with varying degrees of hierarchy, size imbalance, and noise, including LFR and both symmetric and asymmetric multiscale models, and show that it consistently identifies partitions closely aligned with ground truth. Applied to real-world networks in social and neuroscience systems, HCE reveals interpretable modular hierarchies that align with known structural and functional organizations. As a scalable and principled method, HCE offers a general, domain-independent approach to hierarchical community detection with potential applications across biological, social, and technological systems.

# Introduction

A fundamental approach to studying complex systems involves analyzing the interactions among their constituent elements. Often, there is a structural-functional trade-off in how these interactions are measured (Jarrett et al., 2006; Newman, 2003; Sporns, 2013). For instance, we may have access to structural connections, such as roads (Friedrich, 2017) or power grids between cities (Pagani & Aiello, 2013), or acquaintance ties in social networks (Kossinets et al., 2008), or to functional relationships derived from time series data, such as asset prices in financial markets (Mantegna, 1999), gene expression levels in cells (Ali & Alrashid, 2025), climatic variables like temperature and pressure (Donges et al., 2009), or neural activity in the brain (Sarwar et al., 2021; Suárez et al., 2020).

Interactions in real-world systems are generally not randomly distributed but instead exhibit patterns shaped by the system's underlying functional mechanisms. Two recurring organizational motifs observed in such systems are *communities* (Newman, 2003*)* and *hierarchies* (Mariani et al., 2019*)*. Communities are typically defined as strongly coupled subgroups, either densely connected or highly correlated, that are often associated with specific functions within a larger system (Fortunato & Hric, 2016). Hierarchies, on the other hand, can describe multilevel organizational structures, such as top-down or bottom-up regulation in biological systems or institutions (Hooghe & Marks, 2021; Mengistu et al., 2016; Ravasz & Barabási, 2003). Despite their intuitive appeal, there is no universal consensus on how to rigorously define either communities or hierarchies (Fortunato & Hric, 2016; Rosvall et al., 2019). Nevertheless, numerous algorithms have been developed to infer them.

If a system is assumed to have a single partition level, specialized community detection algorithms are employed. These typically involve optimizing a cost function, for example, modularity in the Louvain method (Blondel et al., 2008), the map equation in Infomap (Rosvall et al., 2009), or Bayesian model evidence in stochastic block models (Aicher et al., 2015; Peixoto, 2019). Although these algorithms are effective in systems with a single community level, they are generally ill-suited to systems with inherent hierarchical organization. To address such cases, alternative methods have been proposed, including hierarchical stochastic block models (Heller & Ghahramani, 2005; Schaub et al., 2023), advanced multiresolution consensus clustering (Jeub et al., 2018), link community detection (Ahn et al., 2010; Armas et al., 2024), and hierarchical divisive or agglomerative clustering algorithms (Girvan & Newman, 2002; Sokal & Michener, 1958). These approaches share the common principle that a hierarchically organized system should be representable by a *dendrogram*, a tree-like structure illustrating community merging across different levels of granularity. A major challenge, however, lies in determining which levels of the dendrogram correspond to meaningful community structures.

Our contributions to discovering emergent multiscale structure include the introduction of a novel entropy-based measure, referred to as *hierarchical clustering entropy* (HCE)*,* which quantifies the effective information of community partitions across a dendrogram. This enables the identification of the partition with the most balanced and diverse community size distribution, as measured by Shannon entropy (Shannon, 1948) regularized by the number of communities, capturing a high degree of structural differentiation. Then, we demonstrate that, once such a level is identified, we can apply a block renormalization procedure to uncover additional coarse-grained community levels (Kadanoff, 1966; Villegas et al., 2023; Wilson, 1971). The code used in this study is available in our [GitHub](#) repository.

## Results

**The HCE method**
A dendrogram is a data structure that represents the hierarchical formation of communities in a network, beginning with individual nodes and progressing through successive merges until all nodes are grouped into a single community. Several algorithms, either agglomerative or divisive, can be used to construct such dendrograms. Once a dendrogram is obtained, a natural question arises: which levels, if any, correspond to informative community partitions?

Intuitively, informative levels are those where communities are relatively balanced in size. This is because, in such partitions, identifying the community membership of a randomly selected node requires more information, i.e., more bits or a greater number of questions, than in partitions with a similar number of communities but dominated by a single large group. In information-theoretic terms, these partitions exhibit higher entropy, calculated from the probability distribution of community sizes.

However, this intuition leads to a critical issue: the partition in which every node is a singlet (i.e., its own community) maximizes entropy but provides no insight into the structure of the system. To overcome this, we introduce two key contributions of this work: the notion of an *effective* community, and a corresponding entropy-based measure called hierarchical clustering entropy (HCE).

An effective community is defined as a community from which one node has been randomly removed. This operation ensures that communities of size one (singlets) are excluded from contributing to the entropy, since they contain no clustering information. We then compute the entropy of the probability distribution over these effective communities, yielding a measure of clustering information that favors partitions made of balanced, non-trivial groups.

To account for the removal of one node per community, we multiply the entropy by a normalization factor representing the proportion of nodes retained, relative to the maximum possible, which occurs when all nodes are merged into a single community. The resulting expression defines the hierarchical clustering entropy (HCE; Eq. 1), a novel metric proposed here:

$$\text{HCE}(K_i) = \frac{N - K_i}{N - 1} \sum_{c=1}^{K_i} p_c \log \frac{1}{p_c}, \tag{1}$$

where

$$p_c = \frac{n_c - 1}{N - K_i}. \tag{2}$$

In these equations, $K_i$ is the number of communities at dendrogram level $i$, while $n_c$ and $p_c$ are the size of community $c$ and the probability of sampling a node from that community, respectively. We also refer to $p_c$ as the *effective community fraction*. Note that $p_c = 0$ for any community of size one, which ensures that such trivial communities contribute nothing to the entropy. This design explicitly excludes uninformative partitions from being considered optimal. **Fig. 1a** explains HCE with an example.

With this framework, the partition that maximizes HCE corresponds to the most informative level of community structure in the dendrogram. However, hierarchical

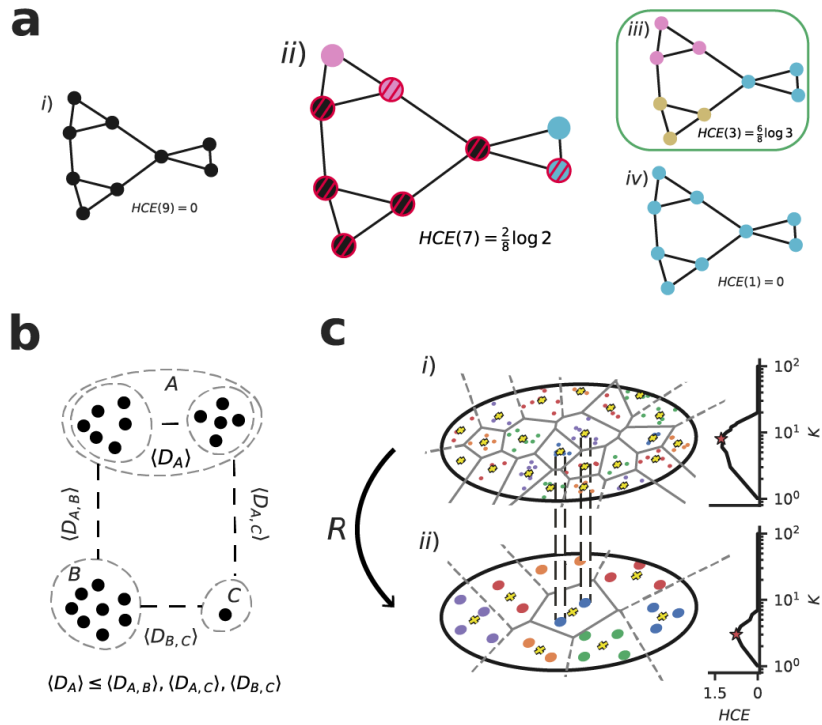

**Fig. 1 | The HCE method. a**, Illustration of HCE computation on a toy network with nine nodes. Nodes are shown in black if they form singlet communities, and in color when part of larger communities. *i*) When each node is its own community, the partition has maximum entropy but zero clustering information, so HCE(9) = 0. *ii*) At a level with seven communities, only two have more than one member, and thus non-zero effective size HCE is computed by randomly removing one node per community; only two effective communities remain, giving HCE(7) = (2/8) $\log 2$. *iii*) A partition into three balanced communities has the highest HCE: HCE(3) = (6/8) $\log 3$. *iv*) At the root level, with one community containing all nodes, HCE is again zero. **b,** In hierarchical clustering algorithms such as UPGMA, the number of communities at each level is determined by a global resolution parameter. At each merge step, the average inter-community distances must be greater than or equal to the merging distance. These inter-cluster distances define a scale at which the system can be renormalized. **c,** Application of the HCE method to randomly distributed points in a 2D ellipse, clustered using the Euclidean distance matrix and UPGMA. *i*) The partition corresponding to the maximum HCE is shown. Each community is represented by a colored region; yellow crosses denote community centroids, and gray lines indicate the Voronoi tessellation based on these centroids. *ii*) After renormalization, each community is replaced by its centroid, yielding a new coarse-grained network from which we can obtain a new partition level of maximum HCE.

organization may involve multiple relevant levels. How can we uncover these additional layers?

To address this, we observe that each dendrogram level corresponds to a specific resolution parameter. For example, in hierarchical agglomerative clustering using the average linkage criterion (UPGMA; Sokal & Michener, 1958; Methods), the average distance between any pair of communities at a given level must be at least as large as the average distance between the communities that merged at that level (see **Fig. 1b**). A similar interpretation applies to other resolution parameters, such as the $\gamma$ in modularity maximization.

We propose that each level in the dendrogram defines a *renormalization scale*, at which each community is treated as a single node. This operation effectively *trims* the dendrogram, removing all structure below the chosen level. The idea is inspired by block renormalization techniques in statistical physics (Villegas et al., 2023; Wilson, 1971), where the system is coarse-grained to study emergent behavior at larger scales.

After renormalization, the dendrogram can be re-analyzed to compute HCE across upstream levels and identify the next optimal partition. This process allows us to uncover a hierarchy of informative scales.

**Fig. 1c** illustrates this concept. In panel **Fig. 1c-*i***, the system is represented at a fine-grained resolution. After identifying an optimal level and applying renormalization, a coarser representation emerges in **Fig. 1c-*ii***. The right panels show the HCE values across levels, with red stars indicating optimal partitions. Thus, applying HCE recursively after renormalization offers a principled way to extract a hierarchy of informative community partitions, rather than a single flat division.

**Benchmarking**

We evaluated the performance of the HCE method on three benchmark models, each designed to probe different aspects of hierarchical structure: (1) the flat hierarchy represented by the Lancichinetti–Fortunato–Radicchi benchmark (LFR; Lancichinetti et al., 2008), which generates networks with a single planted partition (**Fig. 2a**); (2) a symmetric hierarchy generated by the hierarchical nested random graph model (HNRG; Sales-Pardo et al., 2007), which creates multilevel networks where each community recursively splits into an equal number of equally sized subcommunities (**Fig. 2b**); and (3) an asymmetric hierarchy represented by a planted partition hierarchical benchmark (HB; Jeub et al., 2018) that introduces multiple levels of nested structure with heterogeneous community sizes (**Fig. 2c**). Models' descriptions can be found in the Methods section.

For the LFR and HNRG models, dendrograms were constructed using UPGMA applied to cosine distance matrices derived from nodes' connectivity vectors (see Methods). For the HB benchmark, the pronounced imbalance in community sizes across levels motivated the use of dendrograms obtained from the multiresolution consensus clustering (MCC) algorithm (Jeub et al., 2018), which provided a more accurate representation of the hierarchical structure. While UPGMA is computationally efficient, MCC is more demanding, and its performance can be sensitive to network density. UPGMA was implemented via SciPy's `linkage` function (Virtanen et al., 2020) with the `average` method, producing a linkage matrix used to compute HCE and evaluate renormalization levels. This format is equivalent to that produced by MATLAB's `dendrogram` function (*MATLAB*, 2025).

For each parameter setting in each benchmark, we generated 100 network instances and applied the HCE procedure to extract a multiscale sequence of informative partitions from the dendrogram.

**Evaluation framework**

HCE identifies the most informative partition at each level of a dendrogram based on clustering entropy. We denote the first selected level (before any renormalization) as $R_0$ (the finest informative level). After renormalizing the network by collapsing $R_0$'s communities into super-nodes and trim the dendrogram, HCE selects the next level $R_1$, and so on, producing a sequence $R_0, R_1, R_2$, etc.

To evaluate HCE predictions, we compare predicted partitions to the ground-truth using the Adjusted Mutual Information (AMI, Vinh et al., 2009; see Methods). For each benchmark, we quantify how closely the HCE-selected levels align with the planted structure and analyze how this agreement varies across the parameter space.

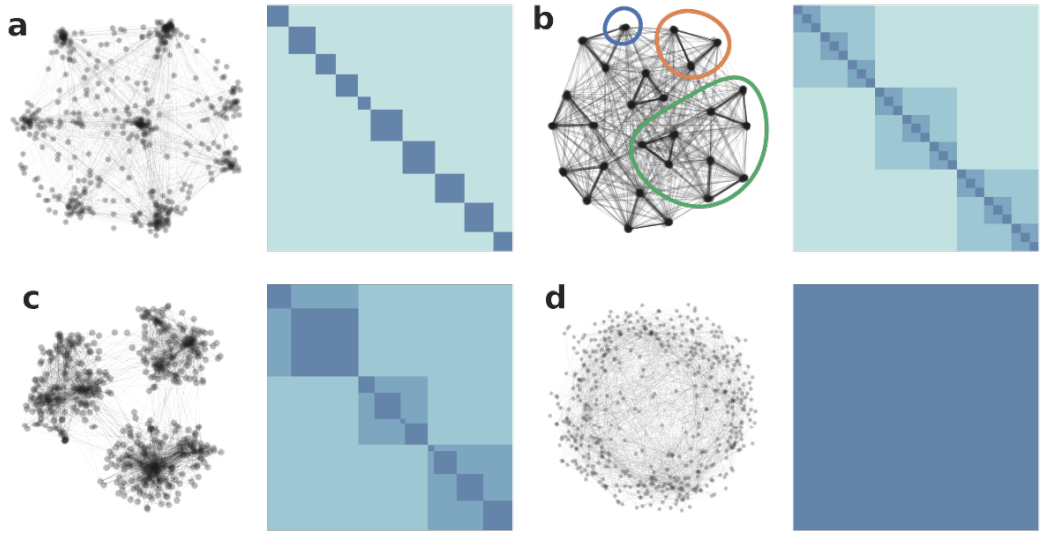

**Fig. 2 | Benchmark.** Representative examples of network architectures used in this study, shown with their corresponding adjacency matrices. Darker blue tones indicate finer-grained community structure, while lighter tones correspond to coarser levels. **a,** LFR benchmark network with a single planted partition (flat hierarchy). **b,** Hierarchical nested random graph (HNRG) model with three symmetric levels of nested communities. **c,** Hierarchical benchmark (HB) network with two asymmetric levels; communities vary in size and number across partition levels. **d,** Erdös–Rényi (ER) random graph, lacking community structure. Network layouts were generated by applying t-SNE (Maaten & Hinton, 2008) to cosine distance matrices derived from node connectivity vectors, embedding nodes in a 2D space. Networks with hierarchical or modular organization form distinct spatial groupings; no such structure is visible in the ER case.

**Flat case: LFR benchmark**

We tested the HCE method on LFR networks with 1,000 nodes across a range of topological mixing parameters $\mu_t$ (the fraction of a node's links that connect outside its planted community). The degree and community size exponents were set to 2 and 1.1, respectively; the average and maximum degrees were 15 and 195; and community sizes ranged from 50 to 200 nodes. As $\mu_t$ increases, the planted community structure becomes less distinct. For each LFR instance, we compared the ground-truth partition to the first four HCE-predicted levels ($R_0$ through $R_3$). As shown in **Fig. 3a**, level $R_1$ consistently yields the highest agreement with the ground-truth and closely approximates the best possible cut in the dendrogram, with a mean absolute AMI difference of $4\% \pm 3\%$ (standard error). This result demonstrates that HCE can successfully recover the planted partition, despite relying solely on the distribution of community sizes and not on edge connectivity patterns. **Fig. 3b** compares HCE to established community detection algorithms (Graph-tool SBM inference (Peixoto, 2019), Louvain, and Infomap), showing comparable performance across the full range of $\mu_t$.

**Symmetric multilevel case: HNRG**

Next, we tested HCE on four-level HNRG networks. These networks are generated by recursively nesting equal-sized subcommunities within each community. The control parameter is the average degree $\langle k \rangle$, which affects the density of within-cluster edges: at low $\langle k \rangle$ clusters are sparsely connected and difficult to distinguish; at high $\langle k \rangle$, internal structure is more clearly resolved.

For evaluation, we compared each HCE-predicted level $R_i$ with the corresponding planted level $L_i$, treating this as a one-shot matching across scales. **Fig. 3c** shows that AMI increases with $\langle k \rangle$ across all levels. Interestingly, the coarser levels ($L_1, L_2, L_3$) exhibit a clear transition from poor to strong recovery. This behavior resembles a phase transition. Empirically, we found that the transition (or critical) points occur when the expected number of connections between communities and their immediate subcommunities at a given partition level exceeds one, i.e., when the connected-component critical threshold is reached (see Methods).

**Asymmetric multilevel case: HB benchmark**
Finally, we evaluated HCE on the HB model with three levels, which introduces asymmetric hierarchical structure with varying community sizes across levels. Here, connection probabilities are defined per level and normalized. The parameters $p_0, p_1, p_2$, and $p_3$ denote the connection probability at each level, from fine to coarse, with $p_3 = 0.05$ fixed as the background connection probability between top-level groups. We varied $p_1$ and $p_2$ over [0,1] in 10 steps, enforcing $p_0 = 1 - p_1 - p_2 - p_3 \geq 0$ and rejecting invalid samples.

To identify partition levels in this benchmark, we used the MCC algorithm, which detects communities across multiple resolution levels. Because this method does not produce a bottom-up dendrogram suitable for our HCE pipeline, we developed a custom procedure (see Methods) to convert its output into a dendrogram compatible with the standard SciPy's linkage or MATLAB's dendrogram format. This approach was selected for its robustness to strong community size imbalance, despite its higher computational cost.

**Fig. 3d** shows the AMI between HCE-predicted partitions and the planted ground truth at each hierarchical level. HCE achieves AMI > 0.5 across most of the parameter space, with the highest performance at the finest and coarsest levels. The intermediate level is more challenging to recover, likely due to stochastic variability and pronounced size imbalance between communities, which make it less well-defined. Insets display the maximum AMI obtained by scanning all partitions in the dendrogram for each level, providing an empirical upper bound. Notably, HCE often matches or closely approaches this bound, demonstrating its ability to detect informative multiscale structure even in challenging asymmetric cases.

**Real-world network examples**

To evaluate the performance and limitations of the HCE method in practical settings, we applied it to three real-world networks: a face-to-face interaction network among high-school students (Mastrandrea et al., 2015; Schaub et al., 2023), a resting-state fMRI average correlation network from the human brain (Hagmann et al., 2008; Jeub et al., 2018), and a spontaneous activity whole-brain larval zebrafish dataset (van der Plas et al., 2023).

**High-school interaction network**
This dataset captures face-to-face interactions among 327 students over five days using wearable proximity sensors (Mastrandrea et al., 2015). Students were grouped into nine classes across four specialization tracks: three focused on biology (BIO), three on mathematics and physics (MP), two on physics and chemistry (PC), and one on physics and engineering (PSI). Network edges represent the frequency of active interactions lasting at least 20 seconds. Because the raw interaction frequency $f$ spans three orders of magnitude (including zero), we applied the transformation $w = \log(1 + f)$ to define edge weights to suppress artifacts due to outliers.

Using the UPGMA algorithm, we computed a dendrogram from the weighted network. **Fig. 4a-*i*** shows the HCE curves for the first three renormalization levels, with peaks at $K = 60$, $K = 9$, and $K = 3$ for levels $R_0$, $R_1$, and $R_2$, respectively.

At $R_1$ (**Fig. 4a-*ii***), the partition into nine communities aligns almost perfectly with the students' class labels, achieving an AMI of approximately 0.98. At $R_2$ (**Fig. 4a-*iii***), the partition merges these into three groups that correspond to broader academic specializations: biology, math and physics, and physics without math emphasis. Although the true grouping of four specializations is also present in the dendrogram (at $K = 4$), the HCE measure favors $K = 3$ since it is more balanced with respect to the number of communities. Specifically, in the renormalized network with nine super-nodes, $K = 3$ yields a larger effective community fraction (Eq. 2) than $K = 4$ (6/8 vs 5/8), resulting in a higher HCE score.

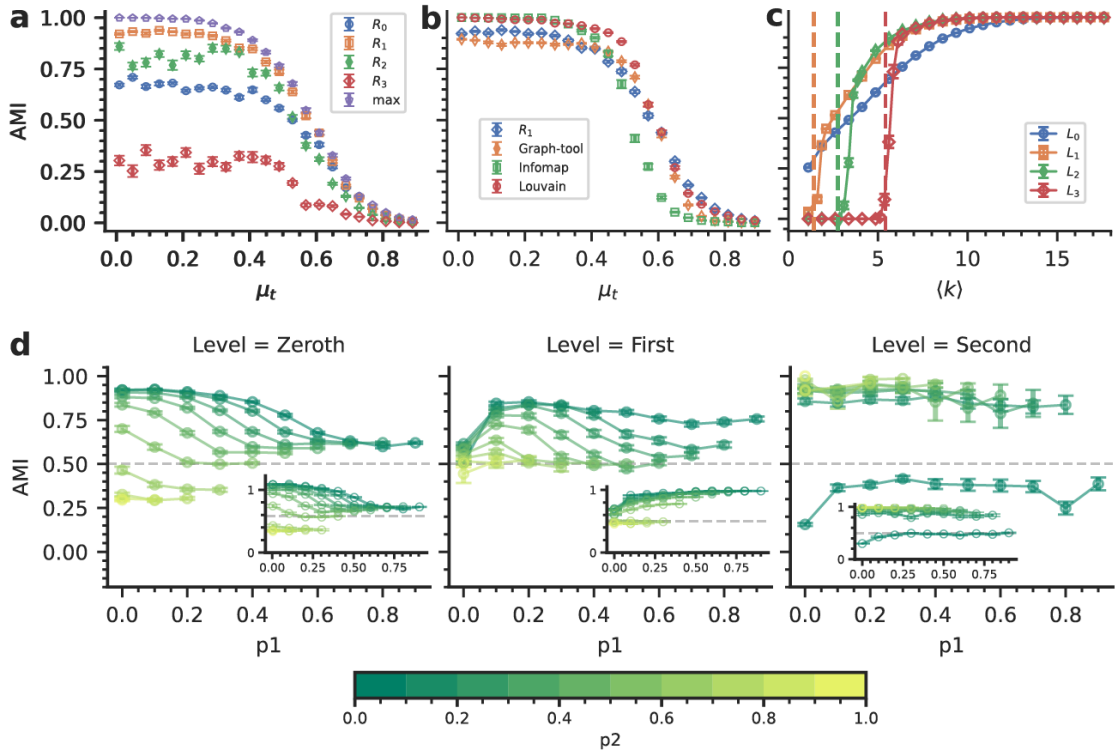

**Fig. 3 | Validation tests. a**, LFR networks were generated to evaluate the performance of HCE in identifying the ground-truth partition. Dendrograms were computed using UPGMA. To assess correspondence, we extracted the first four renormalization levels from the dendrogram and computed the AMI with the planted partition. **b**, The first renormalization level, $R_1$, yields the best agreement with the ground-truth, though its AMI is slightly below the maximum obtained by scanning all possible dendrogram cuts. **c**, HCE was applied to HNRG benchmark networks with four symmetric hierarchical levels. As in **a**, dendrograms were computed using UPGMA. AMI improves with increasing average degree $\langle k \rangle$, but the detectability of each level varies. Vertical dashed lines mark the empirical $\langle k \rangle$ at which the expected number of connections between consecutive levels exceeds the critical connected-components threshold. **d**, Evaluation of HCE on HB networks with three asymmetric levels. Here, communities differ in both number and size across levels. AMI is shown for each level as a function of the first-level connection probability, $p_1$, with color representing the second-level connection probability, $p_2$. The background probability parameter $p_3$ is fixed at 0.05, and $p_0$ is set to ensure normalization. Insets show the best AMI achieved by scanning the dendrogram. HCE closely approaches the optimal in most cases, particularly for the finest and coarsest levels. Each point and bar represent the average and standard error of the mean over 100 network instances.

This example highlights a general feature of HCE: it prioritizes the informativeness of a partition based on balance and size rather than agreement with external labels. While this may occasionally result in partitions that deviate from known categories, HCE still tends to produce close and interpretable groupings, an advantage in unsupervised settings where ground truth is unavailable, such as time-series clustering or exploratory network analysis.

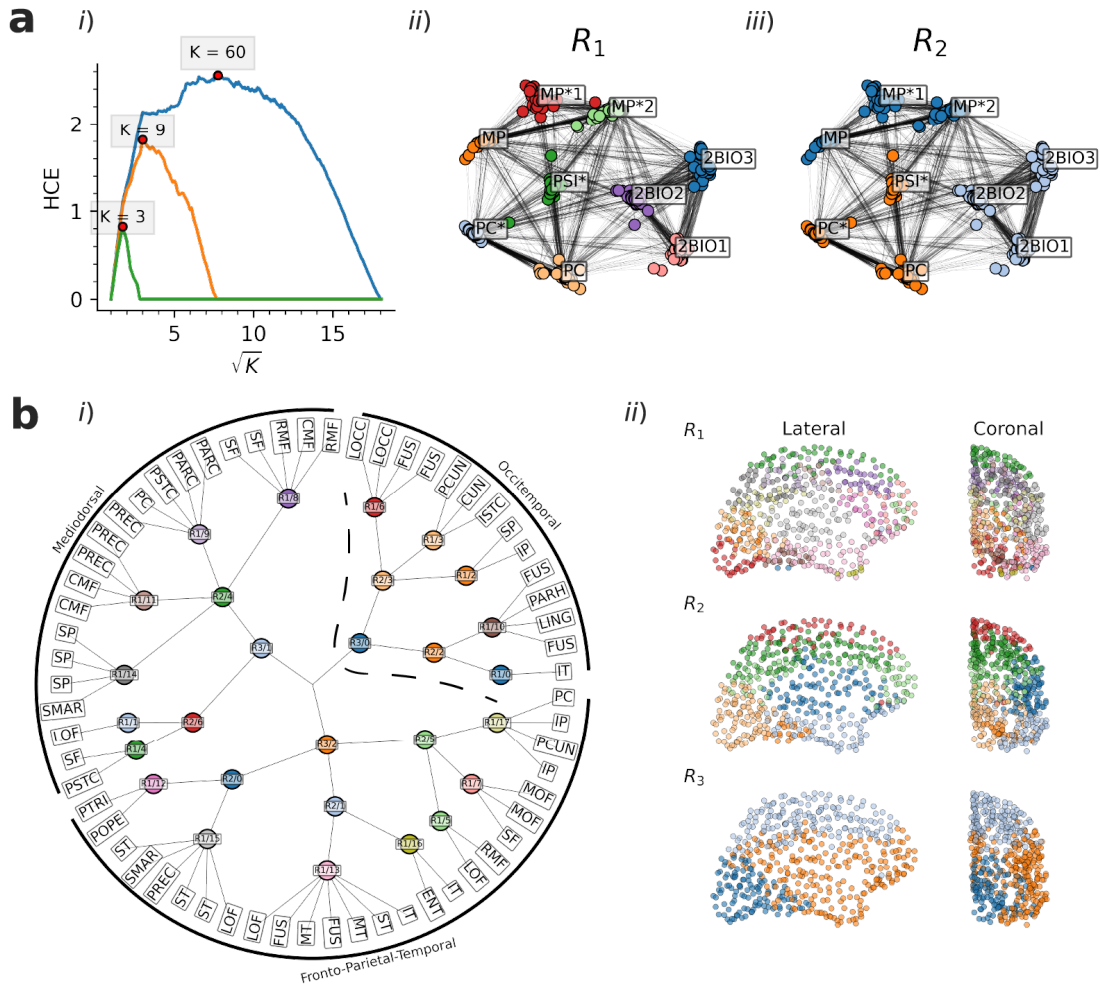

**Fig. 4 | High-school and resting fMRI networks. a**, Results from the high-school face-to-face interaction network. *i*) HCE as a function of the number of communities $K$ for the first three renormalization levels: $R_0$ (blue), $R_1$ (orange), and $R_2$ (green). Peaks at $K = 60$, $K = 9$, and $K = 3$ indicate the most informative partitions at each level. *ii*) The $R_1$ partition ($K = 9$) closely matches the known class structure of the school. *iii*) The $R_2$ partition ($K = 3$) reflects broader academic specialization tracks (biology, math and physics, and physics with no math emphasis). Node positions were obtained via t-SNE on the cosine distance matrix of node connectivity vectors. **b**, Results from the right-hemisphere resting-state fMRI correlation network. *i*) Dendrogram showing the hierarchical organization of ROI clusters across renormalization levels, from $R_0$ (outer ring) to the root ($R_4$, root). Communities at levels $R_1$ through $R_3$ are labeled and colored based on dominant anatomical regions. The dashed line denotes the communities of the visual system branch. *ii*) Spatial projections (lateral and coronal views) of the ROI clusters for $R_1$ (top), $R_2$ (middle), and $R_3$ (bottom). Each color denotes a distinct community at the respective level.

**Resting-state fMRI correlation network**
We also applied the HCE method to a functional brain network constructed from resting-state fMRI data (Hagmann et al., 2008). The network consists of 998 regions of interest (ROIs, see Glossary of terms) spanning both hemispheres. Prior work has shown that functional brain communities tend to be bilaterally symmetric (Hagmann et al., 2008). To avoid redundancy and increase resolution, we analyzed only the right hemisphere, which contains 500 ROIs. Each ROI is annotated with one of 33 anatomical regions based on an established parcellation.

To construct the dendrogram, we converted the average correlation matrix to a distance matrix and applied UPGMA (Methods). Using HCE, we extracted five hierarchical levels ($R_0$ through $R_4$). At $R_0$, the finest level, we identified 67 communities, of which 61 contained more than one ROI. For labeling purposes, we heuristically assigned the dominant anatomical region to each community, as on average 71% of the ROIs in a community shared the same region label (resulting in an AMI of 0.51 between community and anatomical labels). Using these assignments, we built a region-labeled dendrogram showing the hierarchical merging of functional communities (**Fig. 4b-*i***). Colors and labels denote the associated hierarchical levels ($R_1$–$R_3$), with the spatial distribution of ROIs shown in **Fig. 4a-*ii***.

The resulting hierarchy is consistent with known cortical organization. Fine levels ($R_0$, $R_1$) align with anatomical areas. At intermediate level $R_2$, we observe communities that span functional subsystems, such as a light-green cluster covering the dorsolateral prefrontal cortex and precuneus, resembling the default mode network (see Glossary of terms). Other examples include the integration of primary sensory and subparietal regions (e.g., $R_{1/6}$, $R_{1/3}$, $R_{1/2}$; see dashed line) into motion-processing streams ($R_{2/3}$), which subsequently merge with object recognition areas ($R_{2/2}$) at coarser level $R_3$.

Because no ground-truth partition exists for resting-state functional networks, we cannot quantitatively validate the recovered hierarchy. Nonetheless, the multiscale structure revealed by HCE appears functionally meaningful. In future work, such hierarchical representations may aid in understanding how localized functions integrate into large-scale brain systems, and in contrasting these structures across experimental conditions, cognitive states, or clinical populations.

**Hierarchical functional organization in whole-brain larval zebrafish activity**
We analyzed the hierarchical community structure in a publicly available dataset (Fish3) containing spontaneous (i.e., unstimulated) neural activity recorded via two-photon calcium imaging (see Glossary of terms) across the entire larval zebrafish (*Danio rerio*) brain (van der Plas et al., 2023). The dataset consists of time series of relative fluorescence changes ($\Delta F/F$, see Glossary of terms) from approximately 50,000 ROIs, sampled over ~5000 time points spanning 15 minutes. To ensure data quality, we removed traces with unrealistically high or low $\Delta F/F$ values (see Methods), resulting in a final set of ~46000 ROIs. The goal of this analysis is to demonstrate and validate the HCE method, not to exhaustively characterize the larval zebrafish community structure, which would require a consensus analysis across multiple datasets and is beyond the scope of this work.

Community detection seeks to identify groups of neurons (ROIs) with similar activity profiles, potentially reflecting functionally coherent assemblies. A common approach is to compute a pairwise correlation matrix across ROIs, using, for example, Pearson or Spearman correlation, and then cluster this matrix to identify functional communities. However, several challenges arise: (1) the number of ROIs is large ($10^4$–$10^5$) and the resulting correlation matrices are dense, making clustering computationally intensive; (2) many algorithms (e.g., $k$-means) require the number of clusters to be specified in advance; and (3) methods such as modularity optimization or implementing Bayesian inference may fail to identify spatially localized clusters.

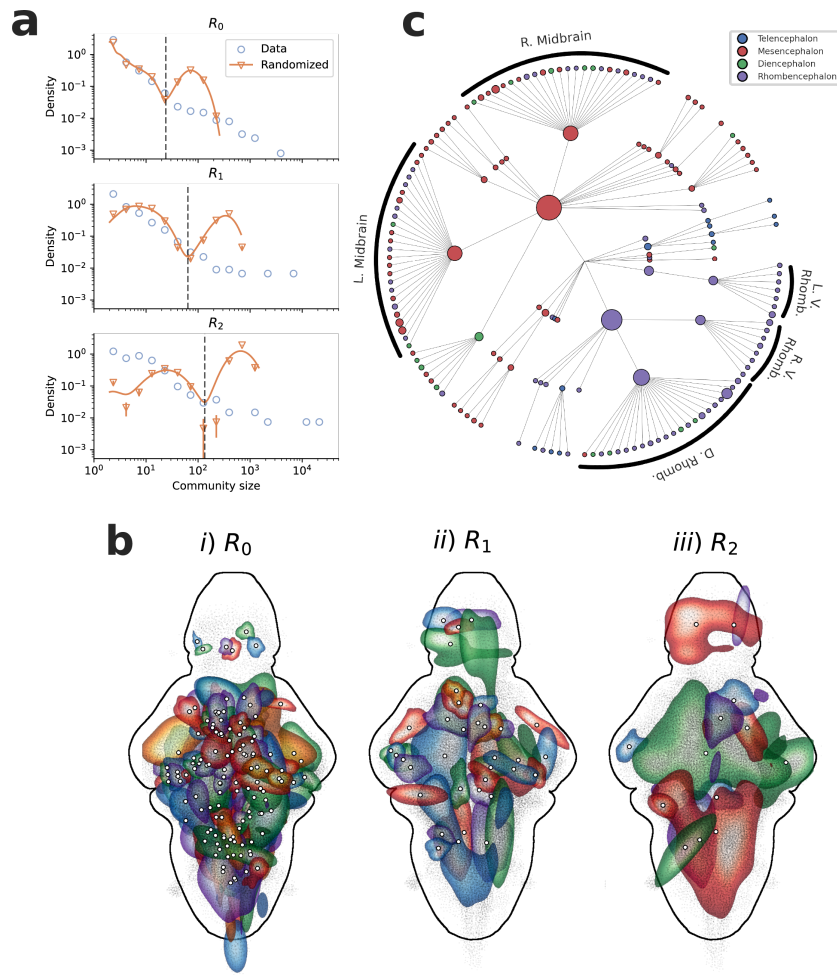

**Fig. 5 | Whole-brain larval zebrafish hierarchical community partition. a**, Community size density distributions across renormalization levels for the larval zebrafish brain data (blue) and the average over 100 null model instances (orange). Bins are spaced logarithmically from $10^0$ to $10^{4.5}$ in 20 steps. Solid curves show Gaussian kernel density estimates for the null model; the dashed line marks the rightmost local minimum, defined as the null-community threshold (NCT). **b**, Communities larger than their NCT across renormalization levels, visualized by fitting a Gaussian kernel to the spatial distribution of their ROIs and displaying only regions above a density threshold (for illustration), exhibit strong spatial contiguity. Individual ROIs are shown as gray points, and community centroids (median coordinates) are marked with white dots. Colors were assigned so that adjacent centroids in the Delaunay triangulation (a planar graph) receive distinct colors, using the greedy coloring algorithm with a largest-first strategy (NetworkX; Hagberg et al., 2008). Mismatches between displayed communities across levels arise from changes in ROI spatial distribution after merging and from the applied kernel density thresholding. **c**, Dendrogram of communities from the first three renormalization levels, with major branches labeled to illustrate anatomical organization. Node sizes are proportional to the fraction of total ROIs in each community. The graph shows a complex partitioning between hemispheres and along the anterior–posterior axis. Abbreviations: D., dorsal; V., ventral; L., left; R., right; Rhomb., rhombencephalon.

As observed in fMRI networks, pairwise correlations in $\Delta F/F$ traces also decay with Euclidean distance (Miri et al., 2011), reflecting anatomical constraints. This motivates the use of UPGMA clustering based on correlation distances, which tends to group spatially adjacent ROIs. We applied HCE to the dendrogram constructed using UPGMA and examined its structure at the first three renormalization levels ($R_0$ to $R_2$).

**Fig. 5a** shows that the distribution of community sizes in the real data approximately follows a power-law, consistent with heterogeneous organization across scales. By contrast, when we applied circular time shifts to scramble the activity traces in null model simulations (see Glossary of terms), the resulting community sizes no longer followed a power-law, instead showing distinct peaks at small and large sizes, patterns likely reflecting constraints introduced by the loss of temporal structure. Notably, the bell-shaped peaks for large communities indicate convergence toward specific sizes absent in the real data.

We quantified this by estimating a *null-community threshold* (NCT) based on the rightmost local minimum point between the two dominant modes in the null distributions, aggregated over 100 null instances and smoothed using a Gaussian kernel. This threshold serves as a baseline to distinguish genuine large-scale communities from those that may arise due to noise or methodological bias.

Using this threshold, we removed small communities and retained only large-scale ones, yielding 124, 35, and 13 communities for $R_0$, $R_1$, and $R_2$, respectively (**Fig. 5b**, panels *i-iii*). These communities show strong spatial contiguity, consistent with anatomical subdivisions. We then constructed a dendrogram (**Fig. 5c**) by tracking their hierarchical merging and colored each community according to the anatomical region with the highest Jaccard overlap (see Methods). At the coarsest level, the brain separates into major branches corresponding to the midbrain (encompassing mesencephalon and diencephalon), the rhombencephalon. The dendrogram also reveals a complex partitioning between hemispheres and along the dorsal–ventral axis.

This case study demonstrates the utility of our approach in biologically realistic systems where modularity optimization is inapplicable. HCE provides a robust, purely structural criterion, based on entropy of community sizes, for identifying informative partition levels from a dendrogram. Furthermore, when spatial constraints are encoded in the correlation structure, the resulting hierarchy reveals how localized assemblies coalesce across scales. This framework offers a promising avenue for analyzing dynamic reconfigurations of brain networks under perturbation, behavioral states, or across experimental conditions.

## Discussion

The identification of meaningful multiscale organization in complex networks remains a central challenge across domains ranging from neuroscience to social systems. While dendrograms offer a natural representation of hierarchical structure, extracting informative levels from them often relies on arbitrary cut thresholds or external supervision. In this work, we introduced hierarchical clustering entropy (HCE), a size-distribution-based metric that selects dendrogram levels by maximizing the effective information content of their community partitions. By recursively applying HCE within a renormalization framework, we obtained a principled sequence of fine-to-coarse partitions, providing an interpretable map of network organization across scales.

Our synthetic benchmark analysis shows that HCE can recover ground-truth partitions in flat, symmetric, and asymmetric hierarchical networks without relying on edge-level connectivity statistics. Even in challenging asymmetric cases with large variations in community size, HCE matched or approached the optimal partition obtainable by exhaustive dendrogram scanning. This robustness stems from its focus on balanced, non-trivial communities, which tend to coincide with structurally coherent modules.

The three real-world applications demonstrate the method's versatility. In the high-school contact network, HCE uncovered both class-level and specialization-level groupings. In the resting-state fMRI network, it revealed functionally meaningful modules spanning multiple anatomical systems. In the larval zebrafish dataset, HCE highlighted spatially contiguous functional assemblies and their hierarchical integration. Together, these examples illustrate how HCE provides a robust, purely structural criterion for identifying informative partition levels from a dendrogram.

There are, however, limitations. HCE assumes that informative partitions are characterized by balance in community size, which may not always hold, particularly in systems where functional relevance is decoupled from size. Additionally, the choice of dendrogram construction method can influence results, as seen in the asymmetric benchmark where UPGMA struggled. Finally, while HCE captures size balance, it does not incorporate connectivity quality directly; integrating both could further improve performance.

Looking forward, HCE's renormalization structure parallels the logic of renormalization group techniques in statistical physics. In neuroscience, the ability to extract reproducible multiscale hierarchies could aid in comparing functional architectures across species, developmental stages, and disease states. More broadly, the method could be applied to any complex system where nested modularity is suspected but not well-defined, providing a general-purpose, interpretable, and scalable tool for hierarchical community detection.

## Methods

### Cosine similarity and distance between node connectivity vectors

The cosine similarity is a measure of the how close two vectors point into the same direction, and its range covers the interval $[-1,1]$. It is defined as

$$\cos \theta_{ij} = \frac{\langle v_I, v_j \rangle}{|v_i||v_j|}, \qquad (3)$$

where $v_i, v_j \in \mathbb{R}^N$, $\mathbb{R}^N$ is the real $N$-dimensional vector space, and $\langle \cdot, \cdot \rangle$ and $|\cdot|$ is the Euclidean dot product and norm, respectively. Then, the cosine distance between the vectors is

$$d_c(\theta_{ij}) = \sqrt{2(1 - \cos \theta_{ij})}.$$

An important feature of the cosine distance is that it is a metric, so it satisfies the triangle inequality.

The Euclidean dot product in Eq. 3 works when vectors belong to a vector space; however, when they denote the connectivity profile of nodes in a graph, the dot product must be adapted accordingly. To understand the reason, see **Extended Data Fig. 1**. In that figure, nodes $i$ and $j$ have the same neighborhood, so their cosine similarity should be equal to one. However, when computed using the standard Euclidean dot product applied to their connectivity vectors (as defined in **Extended Data Fig. 1**) the resulting cosine similarity is

$$\cos \theta_{ij} = \frac{\langle v_I, v_j \rangle}{|v_i||v_j|} = \frac{16}{25} \neq 1.$$

To solve this issue, we define the dot-product on a graph, $\langle \cdot, \cdot \rangle_G$, as

$$\langle v_i, v_j \rangle_G = \sum_{k=1}^{N} v_i(k) v_j(k) (1 - \delta_{ik})(1 - \delta_{jk}) + v_i(j) v_j(i) + v_i(i) v_j(j)$$

Here, $v_i(k) \in \mathbb{R}$ is the connection weight between nodes $i$ and $k$, and $\delta_{ij}$ is the Kronecker delta of indices $i$ and $j$. Using the modified dot product, we get the right result

$$\cos \theta_{ij} = \frac{\langle v_I, v_j \rangle_G}{|v_i||v_j|} = \frac{25}{25} = 1.$$

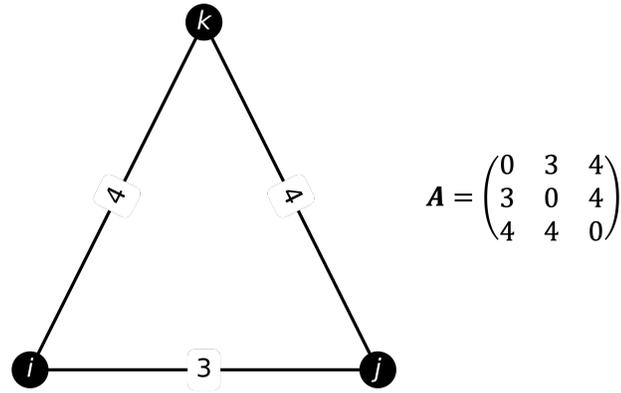

**Extended Data Fig. 1 | A simple toy graph with three nodes.** Nodes $i$ and $j$ share the same neighborhood, illustrating the need for a corrected dot product when computing cosine similarity on connectivity vectors. The weighted adjacency matrix of the graph is also shown; its first two rows correspond to the connectivity vectors of nodes $i$ and $j$, respectively.

In this paper, we measure the cosine distance between nodes in a graph using the dot-product formulation adapted for graphs.

### Correlation distances

The correlation $r$ between two vectors $\boldsymbol{u}$ and $\boldsymbol{v}$ is equivalent to the cosine similarity (Eq. 3) between their $z$-scored versions. Consequently, a correlation measure can be associated with a distance metric, defined as $d = \sqrt{2(1-r)}$, that satisfies the triangle inequality, making it suitable for hierarchical clustering.

### Unweighted pair group method with arithmetic mean (UPGMA)

This is a bottom-up method to build a dendrogram using a distance matrix. It consists in a stepwise process where the nearest two communities merge to form a new community at a new level of the hierarchy. If $A$ and $B$ are two clusters with sizes $|A|$ and $|B|$, then the distance between them in this method is

$$D(A, B) = \frac{1}{|A||B|} \sum_{a \in A} \sum_{b \in B} d(a, b).$$

Therefore, the distance between clusters in the arithmetic mean of the distance between their elements. This is also called average-linkage criterion.

### Multiresolution consensus clustering

This algorithm is designed to compute multiple network partitions across a range of resolution levels (Jeub et al., 2018). It begins by generating an ensemble of partitions through optimization of the Reichardt--Bornholdt formulation (Reichardt & Bornholdt, 2006) of modularity ,

$$Q(\boldsymbol{g}, \gamma) = \sum_{i \neq j} \left[ A_{ij} - \gamma \frac{k_i k_j}{2m} \right] \delta_{g_i g_j},$$

where $A_{ij}$ is the adjacency matrix entry between nodes $i$ and $j$, $k_i$ is the degree of node $i$,

$m$ is the total number of edges, and $k_i k_j / 2m$ corresponds to the Newman-Girvan null model (Newman & Girvan, 2004), which preserves the node degrees. The parameter $\gamma$ controls the resolution of the detected communities, with larger values favoring smaller modules and smaller values favoring larger modules. The vector $\boldsymbol{g}$ specifies the community membership of each node.

The resolution parameter $\gamma$ was sampled using event sampling, a strategy that provides uniform coverage of network scales by monitoring changes in the sign of pairwise modularity contributions. This approach avoids over-sampling high-$\gamma$ values that yield nearly fragmented networks and ensures that the ensemble spans all relevant scales.

From the ensemble of partitions, a co-classification matrix is computed, where each entry represents the fraction of partitions in which a given pair of nodes appears in the same community. Statistical significance of co-classifications is assessed using the *local permutation null model* (Jeub et al., 2018), which preserves the number and sizes of communities while randomizing the membership of one node. This model accounts for variability in cluster sizes and avoids spurious splitting of small clusters.

Consensus clustering is then performed by optimizing a modified modularity function in which the null model is derived from the expected co-classification matrix under the local permutation model. To capture structure at multiple resolutions, a hierarchical consensus procedure is applied recursively: within each consensus cluster, the same procedure is repeated to detect statistically significant subcommunities until no further splits are supported.

**Lancichinetti–Fortunato–Radicchi (LFR) benchmark**

The LFR benchmark (Lancichinetti et al., 2008) generates synthetic networks with a single-level ground-truth community structure and realistic structural features. Such networks are considered to have a flat hierarchical organization. The model produces graphs with power-law degree and community size distributions, closely resembling properties observed in real-world networks. The model parameters control: (1) the number of nodes $N$ and average degree $\langle k \rangle$ (with maximum degree $k_{\max}$), (2) the degree distribution exponent $\tau_1$ and community size distribution exponent $\tau_2$, (3) the minimum and maximum community sizes, (4) and the topological mixing parameter $\mu_t$, which sets the fraction of each node's edges connecting to nodes outside its community.

Nodes are first assigned degrees and community memberships consistent with the chosen distributions. Edges are then placed to satisfy the specified mixing parameter, ensuring tunable levels of community structure from well-separated modules ($\mu_t \approx 0$) to nearly random graphs ($\mu_t \approx 0.5$ or higher).

**Hierarchical nested random graph (HNRG) model**

The HNRG model generates benchmark networks with a predefined multilevel modular organization. Nodes are assigned to nested groups across hierarchical levels, with each node belonging to exactly one group per level and memberships following strict inclusion rules. In this work, the edge probability between two nodes depends only on the finest-grained level at which they share the same community membership.

To construct an HNRG, we first specify the total number of hierarchical levels $L$. The community size at each level $\ell < L$ is denoted by $S_\ell$, with $\ell = 0$ representing the finest-grained level. The top level $\ell = L$ corresponds to the trivial partition in which all nodes belong to a single community. At this level, $S_L$ does not represent an actual community size but rather the number of nodes that do not share membership in any community at finer levels $0 \leq \ell < L$.

Once community sizes across levels are defined, the expected number of intra-community connections for a node at level $\ell$ is

$$k_l = \begin{cases} p_\ell(S_\ell - 1), & \text{if } 0 \leq \ell < L, \\ p_L S_L, & \text{if } \ell = L, \end{cases}$$

where $p_\ell$ is the probability of a connection between two nodes within the same community at level $\ell$. Here, $S_\ell - 1$ accounts for the number of potential connections within a community excluding the node itself.

The connection probabilities are defined as

$$p_\ell = \begin{cases} \dfrac{\rho^\ell}{(1+\rho)^{l+1}} \dfrac{\langle k \rangle}{S_\ell - 1}, & \text{if } 0 \leq \ell < L, \\ \dfrac{\rho^L}{(1+\rho)^L} \dfrac{\langle k \rangle}{S_L}, & \text{if } \ell = L, \end{cases} \qquad (4)$$

where $\langle k \rangle = \sum_{\ell=0}^{L} k_\ell$ is the average degree, and the *cohesiveness*

$$\rho(\ell') = \frac{\sum_{\ell''=\ell'+1}^{L} k_{\ell''}}{k_{\ell'}}$$

is the ratio between the expected number of connections at coarser levels ($\ell'' > \ell'$) and those at level $\ell'$. The model is constructed so that $\rho(\ell') \equiv \rho$ is constant across levels.

Since in the HNRG model the connection probabilities depend only on the depth of the level, it follows that

$$S_\ell = R \, S_{\ell-1} \quad \text{for } 1 \leq \ell < L, \qquad (5)$$

where $R$ is the number of subcommunities at level $\ell - 1$ that merge into a single community at level $\ell$. Note that the total number of nodes is $N = R \, S_{L-1}$. The free parameters are $S_0$ (size of the finest-grained communities), $R$, $L$, $\langle k \rangle$, and $\rho$.

In the HNRG networks studied here, we used $S_0 = 10$, $R = 4$, $L = 4$, and $\rho = 1$. We chose $\rho = 1$ because preliminary analyses indicated that the performance of HCE does not depend significantly on this parameter.

**Hierarchical Benchmark (HB)**
HB generates networks using a multilevel planted partition approach based on a degree-corrected stochastic block model with a power-law degree distribution (exponent 2; minimum degree 5 maximum degree 70; Jeub et al., 2018).

Each network contains $N$ nodes arranged into $L$ nested community levels. The edge distribution is controlled by parameters $p_0, p_1, \ldots, p_L$, with $\sum_{i=0}^{L} p_i = 1$. Following the convention of this study, $p_0$ denotes the fraction of edges within the finest-grained communities, $p_{L-1}$ the fraction within the coarsest-grained communities, and $p_L$ the background edge probability.

The hierarchical structure is generated recursively. First, all nodes start in a single community. Then, each community is split into a random number of subcommunities, drawn from a Poisson distribution with mean 4 and minimum 2. Finally, subcommunity sizes are drawn from a symmetric Dirichlet distribution ($\sigma = 1.5$) and nodes are assigned accordingly. This process is repeated until all $L$ levels are generated, producing progressively finer community structure.

**Adjusted Mutual Information (AMI)**
AMI is a similarity measure between two partitions of a set of $N$ elements or nodes (Vinh et al., 2009). It quantifies the amount of information shared between the two partitions, corrected for the agreement expected by chance.

Given two partitions $U$ and $V$, the mutual information (MI) is defined as

$$\mathrm{MI}(U,V) = \sum_{i=1}^{|U|} \sum_{j=1}^{|V|} \frac{n_{ij}}{N} \log\left(\frac{Nn_{ij}}{a_i b_j}\right),$$

where $n_{ij} = |U_i \cap V_j|$ is the number of elements shared between community $U_i$ in partition $U$ and community $V_j$ in partition $V$, and $a_i = |U_i|$, $b_j = |V_j|$ are the sizes of the corresponding communities.

The adjusted version corrects the mutual information for random agreement by subtracting the expected mutual information $\mathbb{E}[\mathrm{MI}]$. In the SciPy implementation used in this work, $\mathbb{E}[\mathrm{MI}]$ is computed analytically under the assumption that the contingency table entries $n_{ij}$ follow a hypergeometric distribution. The expected mutual information is given by:

$$\mathbb{E}[\mathrm{MI}(U,V)] = \sum_{i=1}^{|U|} \sum_{j=1}^{|V|} \sum_{n=(a_i+b_j-N)^+}^{\min(a_i,b_j)} \frac{n}{N} \log\left(\frac{Nn}{a_i b_j}\right) \Pr(X = n),$$

where $(a_i + b_j - N)^+ = \max(0, a_i + b_j - N)$, and $X$ is a hypergeometric random variable with probability mass function:

$$\Pr(X = n) = \frac{\binom{a_i}{n}\binom{N - a_i}{b_j - n}}{\binom{N}{b_j}}.$$

The AMI is then defined as:

$$\mathrm{AMI}(U,V) = \frac{\mathrm{MI}(U,V) - \mathbb{E}[\mathrm{MI}(U,V)]}{\max\{H(U), H(V)\} - \mathbb{E}[\mathrm{MI}(U,V)]}$$

where $H(U)$ and $H(V)$ denote the entropies of the partitions $U$ and $V$, respectively.

AMI values range from 0 (no agreement beyond chance) to 1 (perfect agreement), with negative values possible when the similarity between partitions is worse than expected by chance. AMI is widely used to assess the accuracy of community detection algorithms by comparing inferred partitions against a known ground-truth.

**Building a bottom-up dendrogram from the MCC algorithm**
The hierarchical consensus procedure from the MCC algorithm produces two outputs: (1) $S_c$, a vector encoding the finest-level consensus partition, where each entry specifies the community assignment of a network node; (2) and **Tree**, a top-down edge list describing merges between communities across successive hierarchical levels, together with their pairwise similarity values. Notably, **Tree** records only community--community merges and does not include direct links between leaf nodes and their communities, so it has less entries than in standard linkage matrices.

To generate a linkage matrix compatible with SciPy's linkage or MATLAB's dendrogram functions, we use a two-stage procedure:

(1) **Completion of the consensus tree:** The partial **Tree** is transformed to match the standard linkage indexing scheme, in which the $N$ original nodes are followed by internal merge nodes numbered from $N + 1$ upward. Cluster indices in **Tree** are renumbered accordingly, and the partition vector $S_c$ is mapped to this new indexing. For each community in $S_c$, edges are added linking it to its constituent network nodes, assigning a similarity value of 1 to represent leaf-level merges. This yields a new matrix **nTree** containing all cluster–cluster and cluster–node relationships.

(2) **Conversion to a linkage matrix: nTree** is traversed in order of decreasing similarity (equivalently, increasing distance). At each similarity level, parent–child relationships are processed: the first two children of each parent are merged into a new internal node, and any additional children are iteratively merged with the most recently created node for that parent. Each merge is recorded in the standard linkage matrix format $[i, j, d]$, where $i$ and $j$ are the indices of the merged nodes or clusters, and $d = 1 -$ similarity is the merge distance. This process continues until all nodes are connected into a single hierarchy.

The resulting linkage matrix encodes the complete hierarchical consensus partition in a format directly usable by the HCE method. For more details on the implementation of this algorithm, see the [GitHub](GitHub) repository.

**Transition points on the HNRG benchmark**

The sharp increases in the AMI curves for HNRG instances (**Fig. 3c**) indicate the presence of critical average degrees, $\langle k \rangle_c(\ell)$, at which the UPGMA-HCE method begins to recover the partition at level $\ell$.

Let $S_\ell$ denote the community size at level $\ell$, with $S_\ell = R\, S_{\ell-1}$ for $1 \leq \ell < L$ (Eq.5), where $R$ is the number of subcommunities at level $\ell - 1$ within each community at level $\ell$. The connection probability between two nodes in the same level-$\ell$ community in an HNRG network with average degree $\langle k \rangle$ is given by $p_\ell(\langle k \rangle)$ (Eq. 4).

For a node in subcommunity $c_{\ell-1} \in c_\ell$, the expected number of connections to other nodes in $c_\ell$ (including its own subcommunity) is

$$k_{\ell-1,\ell}(\langle k \rangle) = p_\ell(\langle k \rangle)(S_\ell - S_{\ell-1}) + p_{\ell-1}(\langle k \rangle)S_{\ell-1}.$$

The first term accounts for connections to other subcommunities within $c_\ell$, and the second term for connections inside $c_{\ell-1}$.

We define the critical average degree for level $\ell$ as the point where each node has, on average, at least one connection within its level-$\ell$ community:

$$k_{\ell-1,\ell}(\langle k \rangle_c) \geq 1.$$

Above this threshold, UPGMA can reliably distinguish level-$\ell$ communities from the background, producing the sharp AMI transitions observed in **Fig. 3c**.

**Calcium trace filtering criteria**

A typical $\Delta F/F$ trace, representing neural activity, is shown in **Extended Data Fig. 2a**. These traces feature a baseline close to zero, with exponential rises and decays at firing events.

Due to the short duration of calcium transients, fluorescence values rarely deviate far from baseline, with peaks typically on the order of 10 standard deviations. A useful measure of this property is the skewness, $\gamma_1$, which quantifies the asymmetry of the distribution:

$$\gamma_1 = \frac{1}{T} \sum_{t=1}^{T} \left( \frac{\Delta F/F(t) - \mu}{\sigma} \right)^3$$

where $T$ is the number of time points, $\mu$ and $\sigma$ are the mean and standard deviation of the trace. For neuronal $\Delta F/F$ signals, we expect a right-skewed distribution ($\gamma_1 > 1$). **Extended Data Fig. 2b** shows the skewness distribution for all ROIs in the Fish3 dataset, with a mean $\langle \gamma_1 \rangle \approx 3$ and a narrow spread. Our first filtering criterion retains only ROIs whose skewness lies within one standard deviation of this distribution (dashed lines in **Extended Data Fig. 2b**).

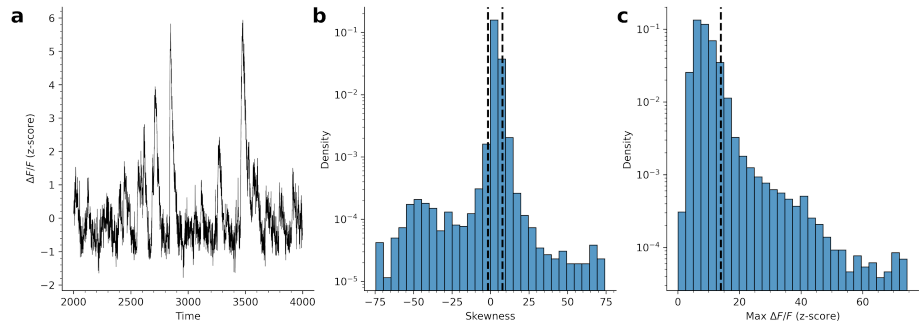

**Extended Data Fig. 2 | Filtering criteria.** (a) Example of a $z$-scored $\Delta F/F$ trace likely representing the activity of a single neuron. (b) Distribution of skewness values across ROIs; dashed lines indicate the mean ± one standard deviation. (c) Distribution of maximum $z$-scored $\Delta F/F$ values; the dashed line marks the mean plus one standard deviation. These criteria were used to filter ROIs in the larval zebrafish analysis.

A second feature considered is the maximum $z$-scored $\Delta F/F$ value of each trace. As shown **Extended Data Fig. 2c**, this distribution is narrowly centered and decays approximately exponentially at higher values. Since our concern is with abnormally large peaks rather than small ones, the second filtering criterion retains only ROIs with a maximum value less than the mean plus one standard deviation of this distribution (dashed line in **Extended Data Fig. 2c**).

**Assigning anatomical regions to clusters using the Jaccard index.** The larval zebrafish dataset (Fish3) was registered to the Z Brain Atlas (Randlett et al., 2015), allowing ROIs to be assigned to one of four major brain regions: telencephalon ($T$), diencephalon ($D$), mesencephalon ($M$), or rhombencephalon ($R$). To assign a community $C$ to a brain region $X \in \{T, D, M, R\}$, we selected the region that maximized the Jaccard index:

$$X^* = \underset{X}{\mathrm{argmax}}\, J(C, X),$$

where

$$J(C,X) = \frac{|C \cap X|}{|C \cup X|}.$$

**Glossary of terms**

Region of Interest (ROI): A manually or algorithmically defined area within an image or image sequence that corresponds to a specific anatomical or functional structure, such as a single neuron, neuronal soma, or dendritic segment, from which a time-varying signal (e.g., fluorescence intensity) is extracted.

Default Mode Network: A functionally connected network of brain regions that shows higher activity during rest or internally focused states and reduced activity during goal-directed or externally focused tasks, as measured by resting-state functional magnetic resonance imaging (fMRI).

Two-photon calcium imaging: A functional imaging approach based on two-photon excitation microscopy which allows the monitoring of changes in calcium concentration inside the cell, a marker of cell activity, in real time and in living brain tissue.

$\Delta F/F$: A normalized measure of relative fluorescence change over time, commonly used in calcium imaging and other fluorescence-based signal recordings to measure neural activity.

Circular shift: It refers to a method of temporally shifting a time series (e.g., fMRI BOLD signal, calcium signals, spike trains) by a fixed number of time steps, where the data that moves past the end is wrapped around to the other end. This preserves the overall shape and temporal autocorrelation of the signal but disrupts its alignment with other signals or events.

**Software packages**
We used Python version 3.11.9 with standard libraries such as Scikit-Learn version 1.5.1, Scipy version 1.15.2, Numpy version 1.23.5, NetworkX version 3.3, Pandas version 2.2.3, Matplotlib version 3.9.1, Seaborn version 0.13.2. In addition, we also used MATLAB version R2025a, and the multiresolution consensus clustering package (Jeub et al., 2018).

**Data availability**
The high-school face-to-face interaction network (Mastrandrea et al., 2015) is available at: https://doi.org/10.1371/journal.pone.0136497.s002
The resting-state fMRI network (Hagmann et al., 2008) is available at: https://figshare.com/articles/dataset/Multiresolution_Consensus_Clustering_in_Networks_-_Network_Data_Sets/5876064/1
The whole-brain larval zebrafish spontaneous activity (Fish3) dataset (van der Plas et al., 2023) is available at: https://gin.g-node.org/vdplasthijs/cRBM_zebrafish_spontaneous_data

**Code availability**
Code used in this research project is available at https://github.com/mrtnzrm2/the_HCE_method.git

**Acknowledgments**
The author would like to thank Prof. Geoffrey Goodhill for his support and guidance during the preparation of this manuscript. Generative AI software was used to assist with text editing under the strict supervision of the author.

**Authors contributions**
J.M.A designed the research, J.M.A wrote all the algorithms and run all the simulations. J.M.A wrote the paper.